\documentclass[11pt]{article}
\usepackage{amsmath,amssymb,color,graphics,epsfig,cite}
\usepackage[table]{xcolor}
\usepackage{amsfonts}
\usepackage{makecell}

\newcommand{\be}{\begin{equation}}
\newcommand{\ee}{\end{equation}}
\newcommand{\bea}{\setlength\arraycolsep{2pt} \begin{eqnarray}}
\newcommand{\eea}{\end{eqnarray}}
\newcommand{\nn}{\nonumber}

\newcommand{\w}[1]{\\[0.#1cm]}
\def\ft#1#2{{\textstyle{\frac{\scriptstyle #1}{\scriptstyle #2} } }}

\def\0{{\sst{(0)}}}
\def\1{{\sst{(1)}}}
\def\2{{\sst{(2)}}}
\def\3{{\sst{(3)}}}
\def\4{{\sst{(4)}}}
\def\5{{\sst{(5)}}}
\def\6{{\sst{(6)}}}
\def\7{{\sst{(7)}}}
\def\8{{\sst{(8)}}}
\def\sst#1{{\scriptscriptstyle #1}}

\def\ep{{\epsilon}}

\def\os{\overline{S}}

\def\omu{\overline{\mu}}
\def\oq{\overline{Q}}
\def\op{\overline{P}}
\def\oa{\overline{\alpha}}
\def\og{\overline{\gamma}}

\def\a{\alpha}
\def\b{\beta}
\def\g{\gamma}
\def\G{\Gamma}
\def\d{\delta}

\def\k{\kappa}
\def\l{\lambda}

\def\m{\mu}
\def\n{\nu}
\def\r{\rho}

\def\t{\tau}

\begin{document}
\begin{center}
{\Large {\bf  Quantum Corrections to Pair Production of Charged Black Holes in de Sitter Space
 }}

\vspace{20pt}

{\large   Yu-Peng Wang, Liang Ma and Yi Pang}

\vspace{10pt}

{\it Center for Joint Quantum Studies and Department of Physics,\\
School of Science, Tianjin University, Tianjin 300350, China }

\vspace{40pt}

\underline{ABSTRACT}
\end{center}
We compute Euclidean action of charged de Sitter black holes in four dimensional gravitational Euler-Heisenberg model. It turns out that the action of a general Euclidean dyonically charged black hole is still controlled by the total entropy contributed by the black hole outer horizon and the cosmological horizon. For smooth configurations, the Euclidean action can be interpreted as the black hole production rate in de Sitter space. We show that
the 4-derivative couplings break the symmetry between the production rate of the purely electric black hole and that of the purely magnetic black hole.  Although electromagnetic duality is no longer a symmetry, it induces a transformation on the 4-derivative couplings, mapping the physical quantities of a purely electric black hole to those of a purely magnetic black hole and vice versa. We also observe that under the same transformation, unitarity constraints on the 4-derivative couplings remain invariant. 

\vfill{\footnotesize   yupengwang@tju.edu.cn \ \ \ liangma@tju.edu.cn \ \ \ pangyi1@tju.edu.cn
}

\thispagestyle{empty}
\pagebreak

\tableofcontents
\addtocontents{toc}{\protect\setcounter{tocdepth}{2}}

\section{Introduction}
Black holes in general relativity have
bridged together gravitation, thermodynamics, and quantum mechanics.  Recent observations of gravitational waves  from binary black holes \cite{LIGOScientific:2016aoc} and neutron stars \cite{LIGOScientific:2017vwq} mergers paved the way
for gravitational wave astronomy looking deeper into the mysterious universe. Undoubtedly, black holes play an increasingly important role both theoretically and experimentally. Therefore, it is
worthwile to understand the cosmological production and evolution of black holes. 

Currently, there are three known mechanisms through which a black hole can be born. The first way is by gravitational collapse \cite{Opp1, Opp2}. A more accurate account on nuclear processes predicts that black holes produced in this channel 
necessarily have a mass greater than 3 solar mass. 
The second way originates from looking for physical processes that could potentially yield Planck size black holes whose Hawking radiation is energetic enough to be detectable. These mini-sized black holes are primordial because their production conditions can only be met in the very early universe when local deviations from homogeneity were sufficiently large \cite{Hawk1}. The third way is analogous to the Schwinger process of pair creation of particles in external field \cite{Sch1}. In the field theory computation, a bounce solution formed by two instantons provides a simple evaluation of the pair production rate. 
Applying the same approach to gravity,  one finds that pair production of black holes can be described by a complex solution  consisting of a Lorentzian section and a Euclidean section. Semiclassically, the wave function corresponding to this process is given by 
\be
\Psi\simeq e^{-I_C}\,,
\ee
where $I_C$ is the action of the complex solution. Thus the probability measure assigned to the process is 
\be
P=|\Psi|^2\simeq e^{-2{\rm Re} [I_C]}\,,
\ee
in which ${\rm Re} [I_C]$ is the action of the Euclidean section equal to the on-shell action of the half-instanton configuration obeying appropriate boundary conditions. 
Gibbons \cite{Gib1} first suggested that in Einstein-Maxwell theory a pair of extremal black holes could be produced in a background magnetic field and the instanton solution responsible for this process is the Euclideanized extremal Ernst solution. Later on this proposal was realized explicitly in \cite{Gar1,Gar2} and generalized to other scenarios \cite{Dow1,Dow2, Ross1, Yi,Brown:1997dm, Mellor:1989gi,Ross2,Garattini:1996ic,BH,BM, Dias} where the external forces pulling apart the virtual black hole pair can be provided by a background electric field, a positive cosmological constant or cosmic strings. 

In this paper, we will compute leading quantum corrections to the production rate of a pair of charged black holes in four-dimensional  Einstein-Maxwell theory extended by 4-derivative interactions which arises from integrating out massive  degrees of freedom charged under the U(1) gauge field. The extended theory is known as the gravitational Euler-Heisenberg model.
Since in the very early and late time, the evolution of our universe is described by a nearly de Sitter space, we would also like to add the cosmological constant in our setup. Thus our work generalizes the scenario studied in \cite{Ross2} to include leading quantum corrections. We will focus on the simple case by assuming spherical symmetry while leaving less symmetric case for future investigations. 
In the standard treatment, instantons describing the pair production of spherically symmetric charged black holes in de Sitter space correspond to smooth configurations of Euclidean charged de Sitter black holes, including the lukewarm, cold and charged Nariai solutions \cite{Ross2}. On the other hand, \cite{Chao:1997osu, Bousso:1998na} suggested that a generic Euclidean charged de Sitter black hole with conical singularities may also describe the pair production of generic charged black holes as long as it is interpreted as constrained instanton. In this latter viewpoint, it is not clear whether the constrained instantons are saddle points or local minima of the action. However, for completeness, we will also compute the on-shell action of a generic Euclidean charged de Sitter black hole, which when specialized to smooth solutions, just yields the  pair production rate of charged de Sitter black holes.

Einstein-Maxwell theory coupled to massive charged particles is a generic subsector of many theoretical models. Integrating out massive degrees of freedom generates  4-derivative 
couplings between the metric and the U(1) field strength at 1-loop level. Within the extended Einstein-Maxwell theory, we obtain generic dyonically charged de Sitter black holes corrected by 4-derivative interactions. Treating the Euclidean charged de Sitter black holes as constrained instantons, we evaluate their on-shell actions. Taking into account contributions from the conical singularity appropriately, we show that the on-shell action of the halved constrained instanton is still given by the simple formula
\be
I_{\footnotesize{\ft12}}=-\ft12 (S_b+S_c)\,,
\ee
where $S_b$ and $S_c$ are the entropy associated with the black hole horizon and the cosmological horizon, extending the result in the 2-derivative theory \cite{Chao:1997osu, Morvan:2022ybp} to include effects from leading 4-derivative corrections. Specicalizing to  smooth configurations such as the lukewarm, the cold and the charged Nariai solutions, we find that electromagnetic duality present at the 2-derivative level \cite{Brown:1997dm, Hawking:1995ap} is broken by 4-derivative interactions. Although electromagnetic duality is no longer a symmetry, it induces a transformation on the 4-derivative couplings, mapping the physical quantities of a purely electric black hole to those of a purely magnetic black hole and vice versa.

This paper is organized as follows. In section 2, we introduce the 4-derivative extended Einstein-Maxwell theory
and present the corrected solution up to first order in 4-derivative couplings. In section 3, we briefly review thermodynamic properties of dyonically charged de Sitter black holes. In section 4, we compute the on-shell action of the half Euclidean black holes. In section 5, we specialize to examples of  smooth configurations which are purely electric or magnetic, exhibiting breaking of electromagnetic duality by 4-derivative couplings. Along the way, we also obtain the corrected charged Nariai solution. In section 6, we discuss how the results from purely electric case can be
related to those from the purely magnetic case via dualization. We conclude with discussions in section 7.

\section{The model}
We shall consider the model below 
\bea
\mathcal{L}&=&\sqrt{-g}\left[\frac{1}{16\pi G}(R-2\Lambda)-\frac1{4g^2}F_{\m\n}F^{\m\n}+\frac{\alpha}{g^2} R_{\mu\nu\rho\sigma}F^{\mu\nu}F^{\rho\sigma}\right.\nn\\
&&\left.\qquad\quad  +\frac{8\pi G}{g^4}\g_1\left(F_{\mu\nu}F^{\mu\nu}\right)^2+\frac{8\pi G}{g^4}\g_2\left(F_{\mu\nu}\widetilde{F}^{\mu\nu}\right)^2\right]\,,
\label{Lag1}
\eea
where $\widetilde{F}_{\mu\nu}=\ft12\epsilon_{\m\n\r\l}F^{\r\l}$. We have adopted a parametrization similar to \cite{CLR}. This theory is known as the gravitational Euler-Heisenberg model.
Later on, we will see that it is $\a,\,\g_1,\,g_2$ which control the magnitude of the corrections relative to the leading term without factors of $G$ or $g$ justifying the legitimacy of the parameterization. We have also performed field redefinitions to get rid of terms proportional to Ricci tensor and U(1) field equation. The effective theory of the form \eqref{Lag1} arises from integrating out massive charged particles. In fact, one can show that the most general parity-invariant 4-derivative interactions can be brought to the form above by field redefinitions \cite{Ma:2020xwi}.   Results from spinor and scalar QED \cite{Drummond:1979pp, Shore:2002gw,Bastianelli:2008cu, Goon:2016une,Alberte:2020bdz} show that
\be 
\a\propto \frac{g^2}{M^2}\,,\quad \g_1\propto {\rm max}[\frac{g^4 M_p^2}{M^4},\,\frac{g^2}{M^2}]\,,\quad \g_2\propto {\rm max}[\frac{g^4 M_p^2}{M_4},\,\frac{g^2}{M^2}]\,,
\ee
where $M$ is the mass of the charged particle. It is clear that $\a,\,\g_1,\,\g_2$ will be at the same order if $g<\frac{M}{M_p}$, otherwise $\a$ is negligible compared to $\g_1,\,\g_2$. 
Also for the 4-derivative interactions to be treated perturbatively, we consider physics below the energy scale set by ${\rm min}[\frac1{\sqrt{\a}},\,\frac1{\sqrt{\g_{1,2}}}]$.
 If the U(1) gauge field couples to multiple massive particles, then 
\be 
\a\propto \sum_i\frac{g_i^2}{M_i^2}\,,\quad \g_1\propto {\rm max}[\sum_i\frac{g_i^4 M_p^2}{M_i^4},\,\sum_i\frac{g_i^2}{M_i^2}]\,,\quad \g_2\propto {\rm max}[\sum_i\frac{g_i^4 M_p^2}{M_i^4},\,\sum_i\frac{g_i^2}{M_i^2}]\,.
\ee

The 2-derivative theory admits   
a dyonically charged de Sitter black hole solution 
\bea
\label{ansatz}
ds^2&=&-h(r)dt^2+\frac{dr^2}{f(r)}+r^2(d\theta^2+\sin^2\theta d\phi^2)\,,\nn\\
h(r)&=&f(r)=1-\frac{2GM}r+\frac{Gg^2q^2}{4\pi r^2}+\frac{Gp^2}{4g^2\pi r^2}-\frac{\Lambda r^2}{3}\,,\nn\\
A_{(1)}&=&\frac{g^2q}{4\pi r}dt+\frac{p}{4\pi}\cos\theta d\phi\,.
\eea
For later convenience, we introduce quantities $M,\,Q,\,P,\,\ell$ with dimension of length
\be
\label{dimen1}
\m=GM\,,\quad Q^2=\frac{Gg^2q^2}{4\pi }\,,\quad P^2=\frac{Gp^2}{4g^2\pi}\,,\quad \Lambda=\frac3{\ell^2}\,,
\ee
in terms of which, the solution becomes
\be
h(r)=f(r)=1-\frac{2\m}r+\frac{Q^2+P^2}{r^2}-\frac{r^2}{\ell^2}\,,\quad A_{(1)}=\frac{gQ}{2\sqrt{\pi G}r}dt+\frac{gP}{2\sqrt{\pi G}}\cos\theta d\phi\,.
\ee
A charged black hole is described by the solution above in the parameter region where $h(r)$ admits 4 real roots. For $M>0$, the structure of $f(r)$ implies that there exist 3 positive roots and 1 negative root. We can thus order them according to $r_n<r_+\le r_b\le r_c$, where $r_c$ stands for the cosmological horizon and $r_b$ is the black hole outer horizon.

The electric and magnetic charges  are defined as
\be
Q_e=-g^{-2}\int_{S^2}\star F=\frac{2\sqrt{\pi}}{g\sqrt{G}}Q=q\,,\quad Q_m=-\int_{S^2} F=\frac{2\sqrt{\pi}g}{\sqrt{G}}P=p\,,
\ee
which satisfy the Dirac quantization condition 
\be
pq=2\pi n\,,\quad  n\in\mathbb{Z}\,.
\ee
Switching on 4-derivative interactions, the solution is modified to 
\be
\label{sol}
h(r)=h_0(r)+\d h(r)\,,\quad f(r)=f_0(r)+\d f(r)\,,\quad A_t(r)=A_{t,0}(r)+\d A_t(r)\,,
\ee
where  $A_t(r)$ is the temporal component of $A_{(1)}$. Functions labeled by subscript ``0" refer to solutions of the 2-derivative theory. Up to first order in 4-derivative couplings the correction terms are given by 
\bea
\d f=&&-\frac{8   \left(P^2-4 Q^2\right)}{r^2 \ell^2} \a -\frac{16 Q^2 }{r^4}\alpha-\frac{4\mu  \left(P^2-7 Q^2\right)}{r^5}\alpha \cr
&&+\frac{8 }{5 r^6}\left( \left(P^2+Q^2\right) \left(P^2-8 Q^2\right)\alpha -2  \left(P^2-Q^2\right)^2 \g_1-8   P^2 Q^2\g_2\right),\cr
\d h=&&\d f+\frac{4 \left(P^2-3 Q^2\right) \left(r^4-r^2 \ell ^2+2 \mu  r \ell^2-\ell ^2 \left(P^2+Q^2\right)\right)}{r^6 \ell ^2}\alpha,\cr
\d A_t=&&\frac{g}{4 \sqrt{\pi\sqrt{G} } }\left[\frac{8   \mu  Q }{r^4}\alpha+\frac{16  Q }{r \ell ^2} \alpha-\frac{4 Q }{5 r^5}\left(P^2 (13 \alpha -16 \g_1 +32 \g_2 )+Q^2 (9 \alpha +16 \g_1 )\right)\right]\,.
\eea

Having obtained the corrected charged black hole solution in de Sitter space, we shall use it to compute on-shell action of Euclidean charged de Sitter black holes in section 4. 

\section{Thermodynamic properties of charged de Sitter black holes}
In this section, we briefly review thermodynamic properties of de Sitter black holes without higher derivative corrections which were first studied in \cite{Gibbons:1977mu} and nicely reviewed in \cite{Bousso:2002fq,Anninos:2012qw}. Here we give a brief summary in our notation for later convenience.
To simplify the discussion, we first introduce the dimensionless parameters
\be
\omu=\m/ \ell\,,\quad \os_{b,c}=S_{b,c}G/\ell^2\,,\quad \oq=Q/\ell\,,\quad \op=P/\ell\,,
\ee
where $S_{b,c}$ denotes the Bekenstein-Hawking entropy associated with the black hole outer horizon and the cosmological horizon. 
Results in \cite{Sekiwa:2006qj,UranoEtal:2009} indicate that one can construct several equalities among thermodynamic quantities. For the black hole mass, there is \cite{Sekiwa:2006qj}
\be
\label{Mformula}
\omu^2=\frac{\os_{b,c}}{4\pi}\left(\frac{\pi \Delta^2}{\os_{b,c}}+1-\frac{\os_{b,c}}{\pi}\right)^2\,,\quad \Delta^2=\oq^2+\op^2\,,
\ee
while for the Hawking temperature associated with the black hole outer horizon and the cosmological horizon,
one can derive \cite{Sekiwa:2006qj} 
\be
{T}_{b,c}=\frac1{8\pi \omu\ell}\left[1-\frac{\pi^2}{\os^2_{b,c}}\Delta^4-2\Delta^2-\frac{4\os_{b,c}}{\pi}+\frac{3\os_{b,c}^2}{\pi^2}\right]\,.
\ee
Setting $T_b=0$ and assuming $\m>0$ leads to
\be
\label{QSrelation}
\Delta^2_*=\frac1{\pi^2}(\pi-3\os_*)\os_*\,,\quad{\rm or}\quad \Delta^2_*=\frac1{\pi^2}(\os_*-\pi)\os_*\,,
\ee
where we use ``*" to label the critical values of $\Delta$ and $\os_b$ at $T_b=0$. 
Substituting the second solution for $\Delta^2_*$ into \eqref{Mformula}, we find $\omu=0$.
In fact this case does not correspond to a degenerate horizon. The only two
real roots of $f(r)=0$ for $\m=0$  are given by 
\be
r_{\pm}=\pm\frac{\ell}{\sqrt{2}}(1+\sqrt{1+4\Delta^2})^{\ft12}\,,
\ee
in which $r_+$ smoothly connects to the cosmological horizon of a pure de Sitter space at $\Delta=0$. In the large charge  limit 
\be
\Delta\gg1\,,\quad r_c\approx \ell \sqrt{\Delta}\,.
\ee
Since the positive real root is connected to the cosmological horizon at zero charge smoothly, it is conceivable that this horizon is akin to the cosmological horizon. 

The interesting case comes with the first solution in \eqref{QSrelation}. Positivity of $\Delta^2_*$
implies that
\be
0\le \os_* \le \frac{\pi}3\,.
\ee
From the first equation in \eqref{QSrelation}, we can solve $\os_*$ in terms of $\Delta^2_*$
\be
\os_{*\pm}=\frac{\pi}6\left(1\pm\sqrt{1-12\Delta^2_*}\right)\,.
\ee
Later on we will see that the two branches of solutions correspond to two extremal situations.
The ``+" one is associated with the large extremal black hole meaning that the black hole outer horizon
meets the cosmological horizon. The `-" one is associated with the small extremal black hole when the black hole outer and inner horizon coincides. The expression above implies an upper bound on $\Delta$ for the existence of extremal black holes,
\be
\Delta^2_*\le \frac1{12}\,.
\ee
Plugging the first equation in \eqref{QSrelation} back to the mass formula \eqref{Mformula},
we can express extremal mass parameter $\omu_*$ in terms of $\os_*$
\be
\label{MSrelation}
\omu_*=\frac{(\pi-2\os_*)\sqrt{\os_*}}{\pi^{3/2}}\,.
\ee
From (17) and (23), we can see that $\os_*= \pi/3$ corresponds to the uncharged Nariai solution.

Substituting \eqref{QSrelation} and \eqref{MSrelation} back to $f(r)$, we can neatly express the four real roots in terms of the critical value of entropy as
\be
\label{4roots}
\{-\frac{\sqrt{\os_*}+\sqrt{\pi-2\os_*}}{\sqrt{\pi}}\,,\quad \frac{\sqrt{\os_*}}{\sqrt{\pi}}
\,,\quad \frac{\sqrt{\os_*}}{\sqrt{\pi}}\,,\quad \frac{\sqrt{\pi-2\os_*}-\sqrt{\os_*}}{\sqrt{\pi}} \}\times\ell\,.
\ee
	\begin{figure}[!ht]
	\centering
	\includegraphics[scale=0.6]{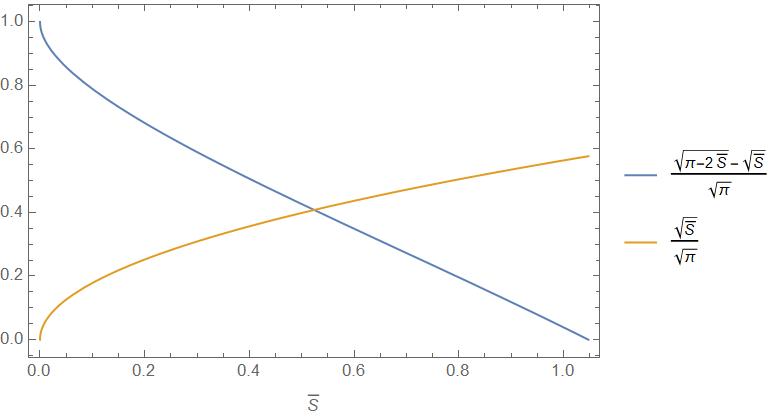}
	\caption{This figure shows the competition between the double root (yellow) representing the location of
the degenerate horizon and the remaining positive root (blue). 
	 }\label{2roots}
	\end{figure}

From Fig. \ref{2roots}, we see that the degenerate root is the largest roots when $\frac{\pi}6<\os_*\le\frac{\pi}3$,
meaning that the outer black hole horizon $r_b$ coincides with the cosmological horizon $r_c$. 
On the other hand, when $0\le\os_*<\frac{\pi}6$, the degenerate horizon is formed from merging the black hole inner horizon $r_{-}$ with the black hole outer horizon $r_b$. In the 2-derivative theory, one can of course replace $\os_*$ by $\pi r^2_*$ where 
$r_*$ is the radius of the degenerate horizon \cite{Ross2}. However, we find that the black hole entropy is a more convenient variable to use when repeating the same analysis for charged de Sitter black holes with 4-derivative corrections. Thus for self-consistency of the paper, we stick to entropy through the whole paper. 
Fig.\ref{2roots} also shows three positive roots coincide at $\os_*=\pi/6$, where
\be
r_-=r_b=r_c=\frac{\ell}{\sqrt6}\,,\quad \Delta^2=\frac{1}{12}\,,\quad \omu=\sqrt{\frac2{27}}\,.
\ee

In the region  $\frac{\pi}6<\os_*\le\frac{\pi}3$ where black hole outer horizon coincides with the cosmological horizon, naively one would think the Euclidean action between $r_b$ and $r_c$ should be 0. However, as shown in \cite{Hawking:1995ap}, one can choose a set of new coordinates in terms of which, the limit $r_b\rightarrow r_c$ is in fact described by 
another smooth geometry with non-vanishing Euclidean action. Here we review derivation of the limiting geometry  in our notation. In order to do so, we first consider 
a small deviation from the coincident limit by tuning the charge and mass slightly away from their extremal values in \eqref{QSrelation} and \eqref{MSrelation}
\be
\omu=\frac{(\pi-2\os_*)\sqrt{\os_*}}{\pi^{3/2}}(1-a^2\epsilon^2)\,,\quad \Delta^2=\frac1{\pi^2}(\pi-3\os_*)\os_*(1+b^2\epsilon^2)\,,
\ee
where $a,\,b$ are arbitrary ${\cal O}(1)$ parameters and $0<\epsilon\ll1$. Accordingly, the four real roots deviate from their values in \eqref{4roots} according to
\bea
\label{limit1}
r_c&=&\ell\sqrt{\frac{\os_*}{\pi}}(1+c_1\epsilon+c_2\epsilon^2+\cdots )\,,\nn\\
r_b&=&\ell\sqrt{\frac{\os_*}{\pi}}(1+d_1\epsilon+d_2\epsilon^2+\cdots )\,,\nn\\
r_-&=& \ell\frac{\sqrt{\pi-2\os_*}-\sqrt{\os_*}}{\sqrt{\pi}}(1+e_1\epsilon+e_2\epsilon^2+\cdots )\,,\nn\\
r_n&=&-\ell\frac{\sqrt{\os_*}+\sqrt{\pi-2\os_*}}{\sqrt{\pi}}(1+f_1\epsilon+f_2\epsilon^2+\cdots )\,,
\eea
where the expansion coefficients are
\bea
\label{2dcoeff}
c_1&=&-d_1=\frac{\sqrt{2a^2(\pi-2\os_*)+b^2(\pi-3\os_*)}}{\sqrt{6\os_*-\pi}}\,,\nn\\
c_2&=&d_2=-\frac{a^2(\pi-2\os_*)^2+2b^2\os_*(\pi-3\os_*)}{(\pi-6\os_*)^2}\,,\nn\\
e_1&=&f_1=0\,,\nn\\
e_2&=&\frac{2a^2(\sqrt{\pi-2\os_*}-\sqrt{\os_*})(\pi-2\os_*)\sqrt{\os_*}+b^2\os_*(\pi-3\os_*)}{2\sqrt{\pi-2\os_*}(\pi(\sqrt{\pi-2\os_*}-5\sqrt{\os_*})+6\os_*(\sqrt{\pi-2\os_*}+\sqrt{\os_*}))}\,,\nn\\
f_2&=&-\frac{2a^2(\sqrt{\pi-2\os_*}+\sqrt{\os_*})(\pi-2\os_*)\sqrt{\os_*}+b^2\os_*(3\os_*-\pi)}{2\sqrt{\pi-2\os_*}(\pi(\sqrt{\pi-2\os_*}+5\sqrt{\os_*})+6\os_*(\sqrt{\pi-2\os_*}-\sqrt{\os_*}))}\,.
\eea
We then perform the coordinate transformation,  
\be
r=\ell\sqrt{\frac{\os_*}{\pi}}(1+c_1\epsilon\cos\chi+c_2\epsilon^2)  \,,\quad \sqrt{\frac{\os_*}{\pi}} c_1\epsilon\tau=\frac{\os_*}{6\os_*-\pi}\ell\psi\,.
\ee
Now plugging (26) and (29) into the solution and taking the limit $\epsilon\rightarrow 0$, we obtain the charged
Nariai solution 
\bea
ds^2&=&\frac{\os_*}{6\os_*-\pi}\ell^2(d\chi^2+\sin^2\chi d\psi^2)+\frac{\os_*}{\pi}\ell^2(d\theta^2+\sin^2\theta d\phi^2)\,,\nn\\
F_{(2)}&=&-\frac{{\rm i}\sqrt{\pi}gQ}{2\sqrt{G}(6\os_*-\pi)}\sin\chi d\chi\wedge d\psi-\frac{gP}{2\sqrt{\pi G}}\sin\theta d\theta\wedge d\phi\,,
\eea
where $Q^2+P^2= \frac{\ell^2}{\pi^2}(\pi-3\os_*)\os_*$. 
This solution generalizes the uncharged Nariai metric \cite{Ginsparg:1982rs} residing at $\os_*=\pi/3$ \footnote{The original Nariai solution is in Lorentzian signature. However, to avoid proliferation of terminology, in this paper, we will refer to its Euclidean version also as the Nariai solution. }. 

\section{Euclidean action of general charged black holes with 4-derivative corrections }
The on-shell action of Euclidean black hole plays an important role in the study of quantum gravity. For instance, the smooth configurations of Euclidean black holes are the instantons describing  pair production of black holes, whose rate is weighted by the action of the instanton. Although generic Euclidean charged de Sitter black holes suffer from conical singularities, we will also compute their on-shell actions following the same strategy as in the instanton case. 
The reason is that it was suggested by Hawking and others \cite{Chao:1997osu, Bousso:1998na } that a generic Euclidean charged de Sitter solution with conical singularities may also describe the pair production of generic charged black holes by interpreting it as a constrained instanton. For the time being, it is not known if the constrained instanton is a saddle point or a local miminum of the action. However, results in this section may become useful in future once the contrained instanton is shown to be a saddle point. Also when specialized to smooth solutions, the result for a generic Euclidean black hole simply yield the probability measure of pair production of charged de Sitter black holes, avoiding repetition for each smooth solution. 

Geometrically, the pair production process is described by a complex solution whose Lorentzian and Euclidean sections are joined at a spacelike hypersurface $\Sigma$ on which the Lorentzian evolution begins. In the bounce approach, the Euclidean section is made from half of an instanton bounded by $\Sigma$ whose second fundamental form vanishes. For pure de Sitter space, $\Sigma$ is topologically $S^3$. 
For non-extremal charged black holes, the surface  $\Sigma$ has topology $S^1\times S^2$, while in the case of extremal charged black holes, the surface  $\Sigma$ has topology $R^1\times S^2$. In this case, there is also a boundary component $B^{\infty}$ representing an internal infinity.

The Euclidean solution  is obtained from the Lorentzian solution via Wick rotation $t=-\rm{i}\tau$. Regularity of the geometry at $r=r_b$ requires the period of $\tau$ be equal to $2\pi/\k_b$ for $\k_b$ being the surface gravity at $r=r_b$. Similarly, regularity at $r=r_c$ requires the period of $\tau$  be equal to $2\pi/|\k_c|$. However, in general, $\k_b\neq |\k_c|$ implying that the Euclidean solution possesses at least one conical singularity. 
	\begin{figure}[!htb]
	\centering
	\includegraphics[scale=0.42]{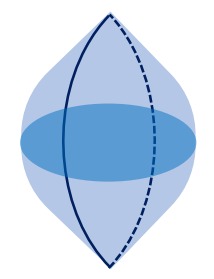}
	\caption{The rugby ball shaped region corresponds to the entire Wick-rotated charged black holes in de Sitter space. Each point in the graph represents a 2-sphere. The rugby tips represent conical singularities at $r=r_b$ and $r=r_c$. The solid curve together with the dashed curve form the boundary of the half-Euclidean charged de Sitter black hole.
	 }\label{halfinstanton}
	\end{figure}
Even though the solution is not entirely smooth, we can still compute its action by carefully taking into account contributions from conical singularities. The recent work \cite{Morvan:2022ybp} suggests that the period of $\tau$ denoted as $\beta$ does not enter the final answer and can take any non-zero value in the intermediate steps. For this reason, we do not need to specify the value for $\b$ in our calculation. However we would like to point out that there is a discontinuity at $\beta\rightarrow\infty$ corresponding to extremal charged black holes. In this case the topology of the instanton is $H^2\times S^2$ instead of $S^2\times S^2$. Thus at the degenerate black hole horizon, the period of $\tau$ is not constrained by any regularity condition and can take any value. The extremal case will be treated separately in the next section.

As shown in Fig.\ref{halfinstanton}, the half-Euclidean solution micking the half-instanton in the pair production process has a boundary along the two arcs at $\tau=0$ and $\tau=\beta/2$ joined at the singularities.  At $\tau=\beta/2$, $r$ goes from $r_b$ to $r_c$ while at $\tau=0$, $r$ runs from $r_c$ to $r_b$, together they form an $S^1$. Thus the topology of the boundary is $S^1\times S^2$. 
The total action of the half-Euclidean solution consists of a bulk term and a boundary term
\be
I_E=I_{\rm bulk}+I_{\rm{bd}}\,,
\ee
where the bulk term  takes the form
\bea
I_{\rm bulk}&=&-\int_0^{\beta/2}d\tau\int_{r_b}^{r_c}dr\int d\Omega_2 \sqrt{g_{E}}\left[\frac{1}{16\pi G}(R-2\Lambda)-\frac1{4g^2}F_{\m\n}F^{\m\n}\right. \nn\\
&&\qquad \quad \left.+\frac{\alpha}{g^2} R_{\m\n\r\l}F^{\m\n}F^{\r\l}+\frac{8\pi G}{g^4}\g_1\left(F_{\m\n}F^{\m\n}\right)^2+\frac{8\pi G}{g^4}\g_2\left(F_{\m\n}\widetilde{F}^{\m\n}\right)^2
\right]\,,
\eea
and the boundary action is 
\bea
\label{4dGH}
I_{\rm{bd}}=-\int_{\Sigma}d^3x\sqrt{h}\left(
\frac{K}{8\pi G}+\frac{\alpha}{g^2} F^{\m\r}F^{\n\l}K_{\m\n}n_\r n_\l+2\frac{\partial {\cal L}}{\partial F_{\m\n}}n_{\m}A_{\n}
\right)\,,
\eea
in which the boundary manifold $\Sigma$ is defined at $\tau=\beta/2$ and $\tau=0$, $n_{\m}$ is the normal vector and $K_{\m\n}$ is the second fundamental form of  $\Sigma$ embedded in the bulk. At $\tau=\beta/2$, $r$ goes from $r_b$ to $r_c$ while at $\tau=0$, $r$ runs from $r_c$ to $r_b$. The first two terms in the boundary
action are the Gibbons-Hawking-York terms needed to ensure the consistency of the path integral in Euclidean quantum gravity \cite{Hawk3}. The third term is added in the boundary action so that one singles out the process with fixed charge instead of potential \cite{Ross2}.

We separate the bulk action in two parts. The first one $I_{s,\,{\rm bulk}}$ receives  only contributions from the smooth geometry. The second one $I_{c,\,{\rm bulk}}$ encodes contributions from conical singularities. After a long calculation, we obtain the bulk action up to the first order in small parameters
\be
\label{smbk}
I_{s,\,{\rm bulk}}=
\frac{\beta}2\left[-\ft12(\psi^e_b-\psi^e_c)Q_e+\ft12(\psi^m_b-\psi^m_c)Q_m+P_{\rm th}(V_c-V_b)+\alpha\Xi+\g_1\Upsilon_1+\g_2\Upsilon_2\right]\,,
\ee
where $\psi^{e}_{b,c}$ stand for the electric potentials on the black hole outer horizon and the cosmological horizon respectively while $\psi^{m}_{b,c}$ denote their magnetic counterparts. $P_{\rm{th}}=-\frac{\Lambda}{8\pi}$ is the pressure term induced by the cosmologial constant. $V_{b,c}$ are the Killing volumes associated with the black hole outer horizon and cosmological horizon. We find that extra quantities $\Xi,\,\Upsilon_1,\,\Upsilon_2$  do not appear in the first law of thermodynamics although they do appear in the on-shell action and the generalized Smarr relation below. Detailed expressions of various quantities introduced here are listed in Appendix A. Using results there we find that up to first order in small parameters the generalized Smarr relation
holds 
\be
\label{msmarr}
0=T_bS_b+T_cS_c+P_{\mathrm{th}}(V_c-V_b)+\frac{1}{2}(\psi^e_b-\psi^e_c)Q_e+\frac12(\psi^m_b-\psi^m_c)Q_m+\alpha\Xi+\g_1\Upsilon_1+\g_2\Upsilon_2\,.
\ee
When higher derivative corrections are turned off, similar Smarr relation without the last three terms  has been obtained in \cite{Dolan:2013ft}.

We now derive contributions from conical singularities. We recall that near the conical singularities Riemann tensor behaves as \cite{Fursaev:1995ef}
\be
\label{cRiem}
R_{\m\n\r\l}=2\pi\epsilon_{b}\left(n_{\m\r}n_{\n\l}-n_{\m\l}n_{\n\r}\right)\delta(r=r_b)
+2\pi\epsilon_{c}\left(n_{\m\r}n_{\n\l}-n_{\m\l}n_{\n\r}\right)\delta(r=r_c)+{\rm regular}\,.
\ee
where $\epsilon_{b,c}=1-\beta T_{b,c}$ and $n_{\m\n}=\sum_{a=1}^2 n^{(a)}_\m n^{(a)}_\n$ is composed by the two orthonormal vectors $n^{(a)}_\m$ orthogonal to the locus of the conical singularity. Substituting \eqref{cRiem} into the action, we obtain
\be
I_{c,\,{\rm bulk}}=-\frac12(\ep_b S_b+\ep_c S_c)\,. 
\ee

In the derivation of $I_{b,\,{\rm bulk}}$ and $I_{c,\,{\rm bulk}}$, we have utilized the fact that 
the solution is invariant under  $\t$-translation, thus omitting the boundary action, the on-shell action of the half-Euclidean black hole equals half of the entire solution.

Using \eqref{msmarr} and (37), 
 we find that the bulk Euclidean action takes the same form as in the 2-derivative case, except that the entropy and the electric potential now encode corrections from the 4-derivative interactions
\be
\label{bulkact}
I_{\rm bulk}=-\frac12(S_b+S_c)-\frac{\b}2(\psi^e_b-\psi^e_c)Q_e\,.
\ee

We now  evaluate the boundary action. Since the gauge potential appears in the boundary action explicitly, we must choose a preferred gauge. We follow the prescription of \cite{Ross2} and set 
\be
A_{(1)}={\rm i}A'_t(r)\t dr+\frac{p}{4\pi}\cos\theta d\phi\,,
\ee
which removes the possible singularity in the electric part of the gauge potential.
A direct computation shows that the magnetic part of the gauge potential does not
contribute to the boundary action in any gauge. Therefore there is no need to fix it into the standard Wu-Yang gauge.  
Substituting the solutions into the boundary action, we obtain 
\be
I_{\rm bd}=\frac{\b}2(\psi^e_b-\psi^e_c)Q_e\,.
\ee
Combining the bulk and boundary actions, we obtain the action of the half Euclidean black hole 
of the simple form 
\be
I_{E}=-\frac12(S_b+S_c)\,.
\label{aofc}
\ee

\section{Smooth instantons and black hole pair producte rates}

 In this section, we focus on smooth Euclidean charged de Sitter black holes. These are instantons describing pair production of charged de Sitter black holes. Based on the result \eqref{aofc} for generic charged de Sitter black holes, we see that the probability measure of pair production of charged black holes in de Sitter space is given by 
\be
P_{\rm BH}=e^{S_b+S_c}\,,
\ee
generalizing the result for uncharged black holes in 2-derivative Einstein gravity with a positive cosmological constant \cite{Morvan:2022ybp}. Given the probability measure of producing a pure de Sitter space \cite{Ginsparg:1982rs}
\be
P_{\rm dS}=e^{\frac{3\pi}{\Lambda G}}\,,
\ee
the rate of producing a charged black hole in de Sitter space is thus determined by the ratio 
\be
\G=\frac{P_{\rm BH}}{P_{\rm dS}}\,. 
\ee

We have mentioned in the previous section that the general result (42) does not apply to the cold solution which will be discussed separately. 
\subsection{The lukewarm solution }
 The first special case is the lukewarm solution with equal temperature on the black hole outer horizon and the cosmological horizon. In our definition of the surface gravity \eqref{tb} and \eqref{tc}, this means  
\be
\k_b+\k_c=0\,,
\ee
so that the conical singularities can be avoided by choosing the period of $\t$ to be 
\be
\b=\frac{1}{T_b}=\frac1{T_c}\,.
\ee
Using general results obtained in the previous section, we find that
since $I_{c,\rm bulk}=0$, the action of the half-lukewarm solution is given by
\be
I^{\rm L}=-\ft12\b(T_bS_b+T_cS_c)=-\ft12(S_b+S_c)\,,
\ee
as expected. Below, in order to see the pair production rate is asymmetric between purely electric black holes and purely magnetic black holes, we express both entropies in terms of the conserved charges. The lukewarm condition implies for the purely electric solution that 
\bea
r_b&=&\frac{1}{2}\left(\ell-\sqrt{\ell(\ell-4\mu)+\delta_{b,1}}\right)+\delta_{b,2}\,,\nn\\
r_c&=&\frac{1}{2}\left(\ell+\sqrt{\ell(\ell-4\mu)+\delta_{c,1}}\right)+\delta_{c,2}\,,\nn\\
Q^2&=&\mu^2+\frac{4 }{5 \ell ^3  }\left( (\alpha +4 \gamma _1)\ell^3-(7 \alpha +8 \gamma _1)(5 \ell -4 \mu ) \mu ^2 \right)\,,
\eea
where the coefficients encoding higher derivative corrections are given by 
\bea
\delta_{b,1}&=&\delta_{c,1}=-\frac{4}{5 \ell ^2 \mu}\left(8 (7 \alpha +8 \gamma _1) \mu^3+4  (\alpha +19 \gamma _1)\ell^2 \mu -10 (5 \alpha +12 \gamma _1)\ell   \mu^2+ (\alpha -16 \gamma _1)\ell^3\right)\,,\nn\\
\delta_{b,2}&=&\delta_{c,2}=\frac{1}{5 \ell ^2 \mu}\left(-4(7 \alpha +8 \gamma _1) \mu^2+4 (2 \alpha +13 \gamma _1)\ell  \mu+(\alpha -16 \gamma _1)\ell^2\right)\,,
\eea
where $\m$ is the parameter introduced in \eqref{dimen1}.
The corrected temperature takes the form  
\be
T_b=\frac{\sqrt{\ell -4 \mu +\delta_T}}{2 \pi  \ell ^{3/2}}\,,\quad \d_T=\frac{4}{5\ell^3}\left(-8(7\a+8\g_1)\m^2+6(\a+4\g_1)\ell\m+3(\a+4\g_1)\ell^2\right)\,,
\ee
and the on-shell action of the half-purely electric lukewarm solution  is
\bea
I_Q^{\text{L}}=-\frac{\pi  \ell  (\ell -2 \mu )}{2 G}+\frac{2 \pi  }{5 \ell ^2 G}\left(4 (7 \alpha +8 \gamma _1) \mu^2- \left(31 \alpha +44 \gamma _1\right)\ell  \mu+4 (2 \alpha +3 \gamma _1)\ell ^2 \right)\,.
\eea
In deriving the formula above, we have evaluated $S_b$ and $S_c$ using the Wald entropy formula. 

For purely magnetic lukewarm solutions, we have 
\bea
r_b&=&\frac{1}{2}\left(\ell-\sqrt{\ell(\ell-4\mu)+\delta_{b,1}}\right)+\delta_{b,2}\,,\nn\\
r_c&=&\frac{1}{2}\left(\ell+\sqrt{\ell(\ell-4\mu)+\delta_{c,1}}\right)+\delta_{c,2}\,,\nn\\
P^2&=&\mu^2+\frac{4 }{5 \ell ^3  }\left( (3\alpha +4 \gamma _1)\ell^3-(\alpha +8 \gamma _1)(5 \ell -4 \mu ) \mu ^2 \right)\,,
\eea
where the coefficients encoding higher derivative corrections are given by 
\bea
\delta_{b,1}&=&\delta_{c,1}=\frac{4}{5 \ell ^2 \mu}\left(-8 ( \alpha +8 \gamma _1) \mu^3-2  (11\alpha +38 \gamma _1)\ell^2 \mu +30 (\alpha +4 \gamma _1)\ell\mu^2+ (7\alpha +16 \gamma _1)\ell^3\right)\,,\nn\\
\delta_{b,2}&=&\delta_{c,2}=-\frac{1}{5 \ell ^2 \mu}\left(4(\alpha +8 \gamma _1) \mu^2-2(7\alpha +26 \gamma _1)\ell  \mu+(7\alpha +16 \gamma _1)\ell^2\right)\,.
\eea
The corrected temperature takes the form  
\be
T_b=\frac{\sqrt{\ell -4 \mu +\delta_T}}{2 \pi  \ell ^{3/2}}\,,\quad \d_T=\frac{4}{5\ell^3}\left(-8(\a+8\g_1)\m^2+6(3\a+4\g_1)\ell\m+3(3\a+4\g_1)\ell^2\right)\,.
\ee
and the on-shell action of the half-purely magnetic lukewarm solution is
\bea
I_P^{\text{L}}=-\frac{\pi  \ell  (\ell -2 \mu )}{2 G}+\frac{2 \pi  }{5 \ell ^2 G}\left(4 (\alpha +8 \gamma _1) \mu^2- \left(13 \alpha +44 \gamma _1\right)\ell  \mu+4 (\alpha +3 \gamma _1)\ell ^2 \right).
\eea
The results above indicate that 
for lukewarm solutions of the same mass , we can distinguish the electric one from the magnetic one from  their
pair production rates since
\bea
I_P^{\text{L}}-I_Q^{\text{L}}=-\frac{4 \pi    }{5 \ell ^2 G}\left(12 \mu ^2+2 \ell ^2-9 \mu  \ell \right)\a\,.
\eea
If comparing the production rate of electric black holes to that of magnetic black holes, obeying $Q=P$, one still finds a difference. This can be readily seen in the limit $\ell\gg \{Q,\,P,\,\m\}$, where the relation between charges and the mass simplifies to
\be
Q^2\approx\m^2+\frac{4 }{5  }(\alpha +4 \gamma _1)\,,\quad P^2\approx\m^2+\frac{4 }{5  }(3\alpha +4 \gamma _1)\,.
\ee
Furthermore, considering the region $\{Q^2,\,P^2\}\gg \{\a,\,\g_1\}$, then at leading order, we have
\be
I_P^{\text{L}}-I_Q^{\text{L}}=-\frac{8\pi    }{5  G}\a\,,\qquad {\rm for}\quad Q=P\,.
\ee

\subsection{The cold solution}
The cold solution corresponds to the case when the black hole inner and outer horizons merge,
{\it i.e.}, $r_-=r_b$. Near the degenerate horizon, the Euclidean geometry approaches $H_2\times S^2$. One thus needs
to add a new boundary $B^{\infty}$ at $r=r_b$. The period of $\tau$ is only fixed by the regularity condition at
the cosmological horizon. A straightforward computation shows that similar to the 2-derivative case \cite{Ross2}, the boundary terms at $B^{\infty}$ do not contribute to the half-instanton action. 
Since the cold solution is a special case of the general instanton studied in the previous section, many results obtained there still apply here. For instance, the bulk action encodes only contributions from the smooth geometry is still given by \eqref{smbk} 
\be
\label{cold}
I_{s,\,{\rm bulk}}=
\frac{\beta_c}2\left[-\ft12(\psi^e_b-\psi^e_c)Q_e+\ft12(\psi^m_b-\psi^m_c)Q_m+P_{\rm th}(V_c-V_b)+\alpha\Xi+\g_1\Upsilon_1+\g_2\Upsilon_2\right]\,,
\ee
where we have specified $\beta$ to be $\beta_c=1/T_c$. Meanwhile, since $T_b=0$, the generalized Smarr relation becomes 
\be
0=T_cS_c+P_{\mathrm{th}}(V_c-V_b)+\frac{1}{2}(\psi^e_b-\psi^e_c)Q_e+\frac12(\psi^m_b-\psi^m_c)Q_m
+\alpha\Xi+\g_1\Upsilon_1+\g_2\Upsilon_2\,.
\ee
The geometry is free of any conical singularity, so contributions from the conical singularity vanish
\be
I_{c,\,{\rm bulk}}=0\,. 
\ee
By virtue of the fact that the boundary action associated with $B^{\infty}$ vanishes, the total boundary action 
takes the same form as (41) with $\b=\b_c$
\be
I_{\rm bd}=\frac{\b_c}2(\psi^e_b-\psi^e_c)Q_e\,.
\ee
Thus using (59)-(62), we find the total action for half of the cold solution with boundary is given  by 
\be
I^{\rm C}_{E}=-\frac12 S_c\,.
\ee
Comparing to the result for a general solution \eqref{aofc}, it seems that
the entropy of the extremal black hole read from the Euclidean action should be 0 \cite{Hawking:1994ii,Hawking:1995fd}.

Concerning physical quantities such as the pair production rate, the ultracold solution
can be viewed as a special case of the cold solution in which the degenerate black hole horizon coincides with the cosmological horizon. The critical value of the entropy in the ultracold limit will be calculated in the next subsection.

\subsection{Charged Nariai solution}
In this section, we will study 4-derivative corrections to the charged Nariai solution arising from the limit $r_b\rightarrow r_c$ of the general charged solution. 
This was done nicely for the 2-derivative Einstein-Maxwell theory in \cite{Hawking:1995ap}. 
To analyze the limit, we will generalize the procedure reviewed in section 3 by first expressing black hole temperature in terms of the black hole entropy and other conserved charges. To simplify the calculation, we define dimensionless 4-derivative couplings as
\be
\oa=\alpha/\ell^2\,,\quad \og_1=\gamma_1/\ell^2\,,\quad \og_2=\gamma_2/\ell^2\,.
\ee
Using results given in Appendix, we have up to first order in small parameters
\bea
T_b&=&T_{b,0}+\d T_b\,,\nn\\
T_{b,0}&=&\frac{1}{8 \pi  \ell  {\omu }}\Big(1-2 (\op^2+\oq^2)-\frac{\pi ^2 (\op^2+\oq^2)^2}{ \os_{b}^2}-\frac{4 \os_{b}}{\pi }+\frac{3  \os_{b}^2}{\pi ^2}\Big)\,,\nn\\
\delta T_b&=&\frac{1}{10 \pi  \ell {\omu } \os_b^4}\Big[
8 \pi ^2  \Big(3 \pi ^2 (\op^2+\oq^2)-\os_b^2+2 \pi  \os_b\Big)\left(  (\op^2-\oq^2)^2\og_1+4  \op^2 \oq^2\og_2 \right)\cr
&&\qquad\qquad+\Big(15 (\op^2-\oq^2 )\os_b^4+\pi ^2 \big(\op^2 (5-8 \oq^2)+14 \op^4-22 \oq^4-5 \oq^2\big) \os_b^2\cr
&&\qquad\qquad +4 \pi ^3 (\op^2+\oq^2) (3 \op^2+\oq^2) \os_b+3 \pi ^4 (\op^2+\oq^2)^2 (\op^2+7 \oq^2)
\Big)\oa
\Big]\,,
\eea
where $\overline{S}_b$ is defined with respect to the corrected black hole entropy via $\overline{S}_b\equiv S_b G/\ell^2$. A similar expression holds for the temperature on the cosmological horizon with $\os_b$ replaced by $\os_c$.  
Likewise, the parameter $\m$ characterizing black hole mass can be written as
\bea
\label{4dmass}
\omu^2&=&\omu_0^2+\delta \omu^2\,,\nn\\
{\omu}_0^2&=&\frac{1}{4 \pi ^3  \os_b}\Big(\pi ^2 (\op^2+\oq^2)+\pi \os_b- \os_b^2\Big)^2\,,\nn\\
\d \omu^2&=&-\frac{\pi ^2 (\op^2+\oq^2)-\os_b^2+\pi  \os_b}{5 \pi \os_b^3}\Big[8 \pi ^2  (\op^2-\oq^2)^2\og_1+32 \pi ^2  \op^2 \oq^2\og_2 \cr
&&+\Big(15 (\op^2-\oq^2) \os_b^2+5 \pi  (\op^2-\oq^2) \os_b+\pi ^2 (\op^2+\oq^2) (\op^2+7 \oq^2)
\Big)\oa
\Big]\,. 
\eea

Based on these formulas, one can then analyze the extremal case by setting $T_b=0$. Compared to the 2-derivative case, the expression for temperature depends on the electric and magnetic charges in a rather
complicated way. To simplify the discussion, we shall focus on purely electric or magnetic de Sitter black holes. In these cases, the coupling $(F_{\mu\nu}\widetilde{F}^{\mu\nu})^2$ does not play any role since its non-vanishing requires the black hole be  dyonically charged.

\subsubsection{Purely electric case}
In this case, the zero temperature condition leads to
\be
\label{cq}
\oq_*^2=\frac{\left(\pi -3 \os_{*}\right) \os_{*}}{\pi ^2}\left(1+
\frac{8 \pi  (\oa +2 \og_1 )-48 (\oa +\og_1 ) \os_{*}}{\os_* }\right)\,,
\ee
from which we immediately notice that
different from the 2-derivative case, the charge is non zero at zero entropy, 
\be
\oq_*^2(\os_*=0)=8(\oa+2\og_1)\,,
\ee
which means if $8(\oa+2\og_1)<0$, the positivity of $\oq_*^2$ would impose a lower bound on $\os_*$. In fact, the validity of the perturbative result already requires that $\os_*\gg{\rm max}[\oa,\,\og_1]$ so that one cannot simply apply \eqref{cq} to $\os_*=0$. 

 Solving $\os_{*}$ in terms of $\oq_*^2$, we find that reality condition for $\os_{*}$ requires the discriminant of quadratic form must be non negative so that
\be
1-48\oa-12\oq_*^2+576(\oa+\og_1)\oq_*^2+576\oa^2\ge 0 \,,
\ee
implying the upper bound for $\oq_*^2$ with corrections from the 4-derivative couplings
\be
\label{cbound}
\oq_*^2\le \frac1{12}+4\og_1\,,
\ee
which is the necessary condition for the existence of  extremal charged de Sitter black holes. 
Upon substituting \eqref{cq} to \eqref{4dmass} with $P=0$, we obtain the fully corrected mass at the extremality 
\be
\label{cmu}
\omu^2_*= \frac{\os_{*} (\pi -2 \os_{*})^2}{\pi ^3}+ \frac{4 (\pi -3 \os_{*})(\pi -2 \os_{*}) (\pi  (9 \oa +16 \og_1 )-6\os_{*} (7 \oa +8 \og_1 ) )}{5 \pi ^3}\,.
\ee
Analogous to the 2-derivative case, using \eqref{cq} and \eqref{cmu}, we can solve for the real roots of $f(r)$ or $h(r)$ which leads to the same answer. With 4-derivative corrections, the double root is modified to 
\be
\label{root1}
\ell \left(\sqrt{\frac{\os_*}{\pi}}+\frac{4  (\pi -3 \os_*)}{\sqrt{\pi \os_*} }\oa \right)(\times 2)\,,
\ee
while the other two single roots are modified according to
\be 
\label{root2}
\ell\left(\frac{\sqrt{\pi-2\os_*}-\sqrt{\os_*}}{\sqrt{\pi}}+\d_1+\d_2\right)\,,\quad -\ell\left(\frac{\sqrt{\pi-2\os_*}+\sqrt{\os_*}}{\sqrt{\pi}}-\d_1+\d_2\right)\,.
\ee
in which the correction terms $\d_1,\,\d_2$ are
\bea
\d_1&=&\frac{(744 \os_*^3-804 \pi  \os_*^2+290 \pi ^2 \os_*-34 \pi^3)\oa+(896 \os^3_*-1016 \pi  \os^2_*+400 \pi ^2 \os_*-56 \pi ^3)\og_1}{5 \sqrt{\pi } (\pi -3 \os_*)^2 \sqrt{\pi -2 \os_*}}\,,\nn\\
\d_2&=&\frac{2(174 \os_*^3-192 \pi  \os_*^2+73 \pi ^2 \os_*-9 \pi ^3)\oa-32 (\pi -2 \os_*) (8 \os_*^2-5 \pi  \os_*+\pi ^2)\og_1}{5 \sqrt{\pi } (\pi -3 \os_*)^2 \sqrt{\os_*}}\,.
\eea
From the discussions carried out in the 2-derivative case, we learn that when $\os_{*}^{\rm crit}<\os_*<\pi/3$, the degenerate root corresponds to $r_b=r_c$ while when $\os_*<\os_{*}^{\rm crit}$ the degenerate horizon is formed by merging $r_-$ with $r_b$. The critical value $\os_*^{\rm crit}$ is defined by the coincidence of the three real roots. 
In order to find $\os_{*}^{\rm crit}$, we need to solve for the triple root $r_-=r_b=r_c$. Built upon the results in the 2-derivative case, demanding $h=h'=h''=0$ or $f=f'=f''=0$ at the triple root yields 
\be
\omu_*=\sqrt{\frac2{27}}+\frac{2\sqrt{6}}5(\oa+4\og_1)\,,\quad \oq_*^2=\frac1{12}+4\og_1\,,\quad
r_-=r_b=r_c=\frac{\ell}{\sqrt6}(1-24\og_1)\,. 
\ee
It is evident that the triple root condition precisely corresponds to the upper bound for $\oq_*$ for the existence of 
the extremal black holes \eqref{cbound}. Substituting the value of the triple root back to \eqref{root1} and \eqref{root2}, we obtain the critical value of $\os_*$
\be
\os_{*}^{\rm crit}=\frac{\pi}6-4\pi(\oa+2\og_1)\,.
\ee

We now proceed to construct the analog of Nariai solution emerging in the limit $r_b\rightarrow r_c$. Similar to the 2-derivative case, we consider small deviations from the extremal case 
\be
\overline{\mu}=
\omu_*(1-a^2\epsilon^2)\,,\quad \overline{Q^2}=\oq_*^2(1+b^2\epsilon^2)\,,
\ee
which leads to the split of the double root
\bea
\label{4dcoeff1}
r_c&=&\ell \left(\sqrt{\frac{\os_*}{\pi}}+\frac{4  (\pi -3 \os_*)}{\sqrt{\pi \os_*} }\oa \right)(1+c_1\epsilon+c_2\epsilon^2+ \cdots ),\nn\\
r_b&=&\ell \left(\sqrt{\frac{\os_*}{\pi}}+\frac{4  (\pi -3 \os_*)}{\sqrt{\pi \os_*} }\oa \right)(1+d_1\epsilon+d_2\epsilon^2+\cdots ),\quad 
\eea
with
\bea
c_1&=&c_{1,0}+c_{1,1},\quad c_2=c_{2,0}+c_{2,1}\,,\nn\\
d_1&=&d_{1,0}+d_{1,1},\quad d_2=d_{2,0}+d_{2,1}\,,
\eea
in which expansion coefficients $c_{1,0},\,c_{2,0},\,d_{1,0},\,d_{2,0}$ are those appearing in the 2-derivative case \eqref{2dcoeff} and coefficients $c_{1,1},\,c_{2,1},\,d_{1,1},\,d_{2,1}$ encode 4-derivative corrections
\bea
\label{4dcoeff2}
c_{1,1}&=&-d_{1,1}=\frac{2 \left(\pi -3 \os_*\right)(\oa A_1+\og_1 B_1)}{5 \os_* \left(6 \os_*-\pi \right)^{3/2} \sqrt{2 a^2 \left(\pi -2 \os_*\right)+b^2 \left(\pi -3 \os_*\right)}}\,,\cr
A_1&=&16a^2(\pi^2-14\pi\os_*+18\os_*^2)+b^2(7\pi^2-108\pi\os_*+216\os_*^2)\,,\cr
 B_1&=&-4(\pi-3\os_*)\left(8a^2(3\pi-8\os_*)+b^2(13\pi-48\os_*)\right)\,,\cr
c_{2,1}&=&d_{2,1}=-\frac{2 (\pi -3 \os_*)(\oa A_2+\og_1 B_2)}{5 \left(\pi -6 \os_*\right)^3 \os_*}\,,\cr
A_2&=&8a^2(3\pi^3-28\pi^2\os_*+188\pi\os^2_*-288\os^3_*)+b^2(13\pi^3-168\pi^2\os_*+1068\pi\os^2_*-1728\os^3_*)\,,\cr
B_2&=&-16(\pi-3\os_*)\left(4a^2(\pi^2-18\pi\os_*+32\os_*^2)+b^2(3\pi^2-44\pi\os_*+96\os^2_*)\right)\,.
\eea
Before taking the limit, we need to perform the coordination transformation from $(r,\,\t)$ to $(\chi,\,\psi)$
\bea
&&r=\ell\left(\sqrt{\frac{\os_*}{\pi}}+\frac{4  (\pi -3 \os_*)}{\sqrt{\pi \os_*} }\oa\right)(1+c_1\epsilon\cos\chi+c_2\epsilon^2) \,,\nn\\
&& \left( \sqrt{\frac{\os_*}{\pi}}+\frac{4  (\pi -3 \os_*)}{\sqrt{\pi \os_*} }\oa\right)c_1\epsilon\tau =\ell\left(\frac{\os_*}{6\os_*-\pi}+\delta_{\tau}\right)\psi\,,
\eea
where coefficients $c_1,\,c_2$ are given in \eqref{4dcoeff1}, (79) and \eqref{4dcoeff2} and $\delta_\tau$ is
\be
\delta_{\tau}=-\frac{2 \left(\pi -3 \os_*\right) \left(\pi  (5 \oa +16 \og_1 )-6 \os_* (\oa +8 \og_1 )\right)}{\left(\pi -6 \os_*\right)^2}\,.
\ee
After taking the limit $\epsilon\rightarrow 0$, the geometry describing the $r_b\rightarrow r_c$ limit now 
takes the form
\bea
ds^2&=&R^2_1( d\chi^2+\sin^2\chi d\psi^2)+R_2^2(d\theta^2+\sin^2\theta d\phi^2)\,,\nn\\
F_{(2)}&=&-\frac{{\rm i}\sqrt{\pi}g\ell\oq_0}{2\sqrt{G}(6\os_*-\pi)}\Psi_{ e}\sin\chi d\chi\wedge d\psi\,,\quad
\oq_0=\frac{\sqrt{\pi -3 \os_*} \sqrt{\os_*}}{\pi }
\eea
where 
\bea
 R^2_1&=&\frac{\ell ^2 \os_*}{6 \os_*-\pi }-\frac{16\ell^2 \left(\pi -3 \os_*\right)^2 (\oa +2 \og_1 )}{\left(\pi -6 \os_*\right)^2}\,,\cr
R^2_2&=&\frac{\ell ^2 \os_*}{\pi }+\frac{8\ell^2\left(\pi -3 \os_*\right)}{\pi } \oa \,,\nn\\
\Psi_{ e}&=&1+\frac{6\left(\os_* (5 \oa +12 \og_1 )-\pi  (\oa +4 \og_1 )\right)}{ \os_*}\,.
\eea
According to \cite{Ross2}, the halved charged Nariai solution describing pair production of charged black holes has a boundary at $\psi=0$ and $\psi=\pi$. Since the charged Nariai solution is smooth, one can compute its bulk action straightforwardly without concerning contributions from the conical singularity. To evaluate the boundary action, we choose the gauge for $A_{(1)}$ \cite{Ross2}
\be
A_{(1)}=\frac{{\rm i}\sqrt{\pi}g\ell\oq_0}{2\sqrt{G}(6\os_*-\pi)}\Psi_{ e}\sin\chi \psi d\chi\,.
\ee
Repeating the process carried out before, we obtain the total action for the halved
charged Nariai solution
\be
I_E^{\rm CN}=-S_*\,,
\ee
where we remind the reader that $S_*=\os_* \ell^2/G$. The charge and mass carried by the solution are given in \eqref{cq} and \eqref{cmu}. 

\subsubsection{Purely magnetic case}

For purely magnetic black holes, the discussion is parallel to that for the purely electric
black hole.  Thus we just present the results below.  
At $T_b=0$, the mass and magnetic charge  can be expressed in terms of the entropy as
\bea
\op_*^2&=&\frac{\left(\pi -3 \os_{*}\right) \os_{*}}{\pi ^2}\left(1+\frac{8\pi (\oa +2 \og_1 )-48 \og_1  \os_{*}}{\os_{*}}\right)\,,\nn\\
\omu_*^2&=&\frac{\os_{*}\left(\pi -2 \os_{*}\right)^2 }{\pi ^3}+\frac{4 \left(\pi -3 \os_{*}\right) \left(\pi -2 \os_{*}\right) \left(\pi  (7 \oa +16 \og_1 )-6 (\oa +8 \og_1 ) \os_{*}\right)}{5 \pi ^3}\,.
\eea

The three horizons coincides at 
\be
r_-=r_b=r_c=\frac{\ell}{\sqrt6}\left(1-12(\oa+2\og_1)\right)\,,
\ee
for special values of the mass and magnetic charge
\be
\omu_*=\sqrt{\frac2{27}}+\frac{2\sqrt{6}}5(3\oa+4\og_1)\,,\quad \op^2_*=\frac1{12}+4(\oa+\og_1)\,.
\ee
Surprisingly, the critical value of the black hole entropy at which three horizons coincide is still given by
\be
\os_*^{\rm crit}=\os_{*}^{\rm crit}=\frac{\pi}6-4\pi(\oa+2\og_1)\,.
\ee

The double root corresponding to $r_b=r_c$ is 
\be
\ell \sqrt{\frac{\os_*}{\pi}}\times (2)\,,
\ee
where the higher derivative corrections are all encoded in the total entropy. Repeating the procedure performed in the electric case,  we obtain the generalized Nariai geometry in the limit $r_b\rightarrow r_c$ for the purely magnetic de Sitter black hole below 
\bea
ds^2&=&R^2_1( d\chi^2+\sin^2\chi d\psi^2)+R_2^2(d\theta^2+\sin^2\theta d\phi^2)\,,\nn\\
F_{(2)}&=&\frac{g\ell\op_0}{2\sqrt{\pi G}}\Psi_{ m}\cos\theta d\theta\wedge d\phi\,,\quad
\op_0=\frac{\sqrt{\pi -3 \os_*} \sqrt{\os_*}}{\pi }
\eea
where 
\bea
R^2_1&=&\frac{\ell ^2 \os_*}{6 \os_*-\pi }-\frac{8\ell^2\left(\pi -3 \os_*\right) \left(\pi  (\oa +4 \og_* )-12 \og_1  \os_*\right)}{\left(\pi -6 \os_*\right)^2}\,,\quad R_2^2=\frac{\ell ^2 \os_*}{\pi },\cr
\Psi_{m}&=&  1+\frac{4\pi  (\oa +2 \og_1 )-24\og_1  \os_*}{\os_*} \,.
\eea
In the magnetic case, the boundary term does not contribute and the on-shell action of the half charged Nariai solution is obtained by simply evaluating the bulk action on the solution above. The result is given by 
\bea
I_E^{\rm CN}=-S_*\,.
\eea

\section{Electromagnetic duality}
It is well known that Maxwell equations with source terms are asymmetric between electricity and magnetism. This
fact implies that after integrating out the charged source, the resulting effective action would also break electromagnetic duality, which is confirmed by our results in the previous section. However, we also notice that
many gauge invariant quantities and equalities in the purely electric case can be related to those in the purely magnetic case by a simple map \footnote{(76) and (90) indicate that the entropy 
of the triple degenerate horizon is invariant under electromagnetic duality.}
\be 
\a^{e}=-\a^{m}\,,\quad \g_1^{e}=\g_1^{ m}+\a^{ m}\,,
\ee
together with interchanging $Q$ and $P$. 
Interestingly, one can check that the map itself is invariant under electromagnetic duality, in other words, the transformation above and identity form a $\mathbb{Z}_2$ group. To understand the origin of this map, we follow the standard steps of dualizing the electric Lagrangian based on $A_{\m}$. We start from the Lagrangian below
\bea
\mathcal{L}&=&\sqrt{-g}\left[\frac{1}{16\pi G}(R-2\Lambda)-\frac1{4g^2}F_{\m\n}F^{\m\n}+\frac{\alpha}{g^2} R_{\mu\nu\rho\sigma}F^{\mu\nu}F^{\rho\sigma}+\frac{8\pi G}{g^4}\g_1\left(F_{\mu\nu}F^{\mu\nu}\right)^2\right.\nn\\
&&\left.\qquad \quad +\frac{8\pi G}{g^4}\g_2\left(F_{\mu\nu}\widetilde{F}^{\mu\nu}\right)^2+\ft12\epsilon^{\m\n\r\l}\partial_\m A'_{\n}F_{\r\l}\right]\,,
\eea
where $F_{\m\n}$ is treated as an independent variable. After integrating out $F_{\m\n}$ we obtain a new Lagrangian based on the magnetic vector potential $A'_\m$ and the magnetic coupling $g'=1/g$. Via a further redefinition of the metric
\be
g^{\m\n}\rightarrow g^{\m\n}+\frac{16\pi G }{g^2}\a (g^{\m\n} F_{\r\l}F^{\r\l}-4F_{\m\r}F_{\n}{}^{\r})\,,
\ee
we arrive at the dual Lagrangian in a similar form to the origin Lagrangian 
\bea
\mathcal{L'}&=&\sqrt{-g}\left[\frac{1}{16\pi G}(R-2\Lambda)-\frac1{4g'^2}F'_{\m\n}F^{'\m\n}+\frac{\a'}{g'^2} R_{\mu\nu\rho\sigma}F^{'\mu\nu}F^{'\rho\sigma} \right.\nn\\
&&\left.\qquad+\frac{8\pi G}{g^4}\g'_1\left(F'_{\mu\nu}F^{'\mu\nu}\right)^2+\frac{8\pi G}{g^4}\g'_2\left(F'_{\mu\nu}\widetilde{F}^{'\mu\nu}\right)^2\right]\,,
\eea
where the new effective couplings are related to the old ones by
\be
\label{dualtrans}
\a'=-\a\,,\quad \g_1'=\g_1+\a\,,\quad \g_2'=\g_2+\a\,,
\ee
which is precisely (95).
In deriving the results above, we have used the identity 
\be
(F_{\m\n}\widetilde{F}^{\m\n})^2=4 F^{\m}{}_{\a}F^{\n\a}F_{\m\a}F_{\n}{}^{\a} -2(F_{\m\n}F^{\m\n})^2\,. 
\ee
Curiosity guides us to check if the unitarity constraints on the 4-derivative couplings are invariant under the transformation above. Indeed,  available  unitarity bounds on the 4-derivative couplings are  
invariant combinations under the transformation \eqref{dualtrans}. To be specific, parameters $(a_1',\,a'_2,\,b_3')$ used in \cite{Alberte:2020bdz} are related to ours
by 
\be 
\frac{a'_1}{m^4}=8\pi G\g_1+4\pi G \a\,,\quad \frac{a'_2}{m^4}=8\pi G\g_2+4\pi G \a\,,\quad b_3'=\a m^2\,.
\ee
Assuming the  gravitational $t$-channel pole can be removed, the unitarity bounds obtained in \cite{Cheung:2014ega,Bellazzini:2019xts,Alberte:2020bdz}  are expressed as
\bea
a'_1+a'_2>0&\Rightarrow& \g_1+\g_2+\a>0\,,\nn\\
 4\frac{a'_1}{m^4}>8\pi G\frac{|b_3|}{m^2}&\Rightarrow&(2\g_1+\a)>\ft12|\a|\,,\nn\\
a_2'>0&\Rightarrow&(2\g_2+\a)>0\,,
\eea
which are obviously invariant under \eqref{dualtrans}. In fact, coefficients $a_1'$ and $a_2'$ are already invariant under \eqref{dualtrans} by themselves while the coefficient $b_3'$ flips sign. Based on 
methods developed recently \cite{Zhang:2020jyn,Tokuda:2020mlf,Caron-Huot:2021rmr,Du:2021byy,Chowdhury:2021ynh,Li:2021lpe}, effects from the gravitational $t$-channel pole \cite{Henriksson:2022oeu}
have been considered in a more appropriate way. Interestingly, the state of art bounds on the 4-derivative couplings are still invariant under \eqref{dualtrans}. Of course, bounds  expressed in terms of $a_1'$ and $a_2'$ are trivially invariant. However, the $\mathbb{Z}_2$ action on $b_3'$ explains why the bound depends on $|b_3'|$ or even power of $b_3'$ (See (3.66) in \cite{Henriksson:2022oeu}).

\section{Summary and discussion}
In this paper, we studied corrections to the pair production of charged de Sitter black holes in gravitational Euler-Heisenberg model. We computed
the on-shell action of a generic half Euclidean charged de Sitter black hole and obtain the simple result 
\be
I_{\ft12 }=-\ft12(S_b+S_c)\,.
\ee
We have extended the Smarr formulas for de Sitter black holes to include 4-derivative corrections. 
We then specialized to examples of purely electric or magnetic smooth configurations including the lukewarm solution, the cold solution and the charged Nariai solution. These solutions describe the pair production of charged de Sitter black holes with a rate 

\be
\Gamma=e^{S_b+S_c-\frac{3\pi}{\Lambda G}}\,.
\ee

 We showed that the electromagnetic duality symmetry is broken by generic 4-derivative couplings. However, one can still utilize the duality transformation to relate the results in the purely electric case to those in the purely magnetic case. Along the way, we also observe that under the same transformation, unitarity constraints on the 4-derivative couplings remain invariant. 
This observation should hold even for couplings with more than 4 derivatives. We thus suggest that combinations of Wilson coefficients invariant under the duality transformation are more meaningful and preferred. 

 To fully understand the role of  charged constrained  instantons played in the pair production of charged de Sitter black holes, one should compute the one loop corrections to the partition function \cite{Volkov:2000ih}. This requires understanding the spectrum or heat kernel of perturbations in the background of constrained instanton which is technically more involved than in the smooth background due to the presence of conical singularities. Only when the constrained instanton is shown to be a saddle point in the quantum gravity path integral, we can be sure that the constrained instanton indeed describes
the pair production of charged black holes in de Sitter space.

\section*{Acknowledgment}
We are grateful to Gary Gibbons, Raphael Bousso and Shuang-Yong Zhou for useful communications.
The work is supported in part by the National Natural Science Foundation of China (NSFC) under grant No.~12175164.

\appendix
\section{4-derivative corrected thermodynamic quantities}
We first give definitions for various thermodynamic quantities that appeared in section 3. 
Strictly speaking, results presented in this section are valid only for non-extremal black holes. However, one can check that equalities derived from them such as \eqref{msmarr} are still applicable in the extremal case. 

The temperature on the black hole horizon the black hole horizon is defined through the surface gravity
\be
T_b=\frac{\k_b}{2\pi}\,,\quad \k_b=\frac12 \sqrt{\frac{f}h} h'\big{|}_{r=r_b}\,.
\label{tb}
\ee
For cosmological horizon, the temperature is given by 
\be
T_c=-\frac{\k_c}{2\pi}\,,\quad \k_c=\frac12 \sqrt{\frac{f}h} h'\big{|}_{r=r_c}\,.
\label{tc}
\ee
Compared to the temperature on the black hole horizon, the extra minus sign is needed to render the temperature on the cosmological horizon positive.
Using the induction tensor (2-form)
\be
{\cal M}_{\m\n}=2\frac{\partial {\cal L}}{\partial F^{\m\n}}\,,
\ee
we can obtain the electric charge and the magnetic potential as
\be
Q_{e}=\int_{S^2}\star{\cal M}\,,\quad d\psi^{m}=i_{\partial_t}\star{\cal M}\,.
\ee
The magnetic charge and the electric potential are calculated from 
\be
Q_{m}=-\int_{S^2}F\,,\quad \psi^{e}=i_{\partial_t}A_{(1)}\,. 
\ee
The Killing volumes associated with the black hole horizon and the cosmological horizon are computed from
\be
V_{b,c}=\int_0^{r_{b,c}}dr \int d\Omega_2 |\partial_t| \frac{r^2}{\sqrt{f}}\,.
\ee
To present the thermodynamic quantities for the corrected solution, we separate the final results into the one
from 2-derivative theory and the correction terms. Results in the 2-derivative theory carry the subscript ``0". 
Only the electric and magnetic charges do not receive corrections. For simplicity, we present the expressions for
quantities defined with respect to the black horizon explicitly. Quantities defined with respect to the cosmological horizon are obtained by replacing $r_{b,0}$ with $r_{c,0}$ except that for temperature, an extra minus sign should be added for the aforementioned reason .  We thus have  
\bea
T_b&=&T_{b,0}+\delta T_b\,,\quad S_b=S_{b,0}+\delta S_b\,,\quad \psi^e_{b}=\psi^e_{b,0}+\delta\psi^e_{b}\,,\nn\\
\psi^m_{b}&=&\psi^m_{b,0}+\delta\psi^m_{b}\,,\quad Q_e=q\,,\quad Q_m=p\,,\quad V_{b}=V_{b,0}+\delta V_b\,. 
\eea
in which
\bea
T_{b,0}&=&-\frac{P^2+Q^2-r_{b,0}^2}{4\pi r_{b,0}^3}-\frac{3r_{b,0}}{4\pi\ell^2}\,,\quad S_{b,0}=\frac{\pi }{G}r^2_{b,0}\,,\quad \psi^e_{b,0}=\frac{gQ}{2 \sqrt{\pi G} r_{b,0}}\nn\\
\psi^m_{b,0}&=&\frac{P}{2g\sqrt{\pi G}r_{b,0} }\,,\quad V_{b,0}=\frac{4 \pi }{3 G}r^3_{b,0}.
\label{Einstein thermodynamic}
\eea
where $r_{b,0}$ is the unmodified radius of black hole outer horizon. 

Corrections to the thermodynamic quantities associated with the non degenerate black hole horizon take the form
\bea
&&\delta T_{b}=-\frac{1}{20 \pi ^2  r_{b,0}^{10} T_{b,0}}\Big[
 \Big(45 (P^2-Q^2) r_{b,0}^8\ell^{-4}+ 5 (Q^2-P^2) r_{b,0}^4-2 (P^2+Q^2) (P^2+7 Q^2) r_{b,0}^2
\cr 
&&\qquad+(P^2+Q^2)^2(P^2+7 Q^2)+6  r_{b,0}^4\ell^{-2} (5 (P^2-Q^2) r_{b,0}^2+12 P^2 Q^2-P^4+13 Q^4)\Big)\alpha\cr
&&\qquad+  8 \big((P^2+Q^2-2 r_{b,0}^2 )+9 r_{b,0}^4\ell^{-2} \big)\big(\g_1  (P^2-Q^2 )^2+4 \g_2 P^2 Q^2\big)
\Big]\,,
\nn\w2
&&\delta S_{b}=\frac{1}{5 G  r^5 T_{b,0}}\Big[
  8 (\g_1  (P^2-Q^2)^2+4 \g_2  P^2 Q^2)+ (5 (P^2-Q^2)  r_{b,0}^2\ell^{-2} (3 r_{b,0}^2+\ell ^2)\cr
&&\qquad\quad+ (P^2+Q^2)(P^2+7 Q^2))\alpha
\Big]\,,\nn\w2
&&\delta \psi^e_b=-\frac{gQ}{20\sqrt{\pi^3 G} r^8_{b,0}T_{b,0}}\Big[
\left(\ell ^{-2} \left(6 P^2-42 Q^2-5 \ell ^2\right) r_{b,0}^4+2  \left(9 P^2+7 Q^2\right) r_{b,0}^2+45 r_{b,0}^8
\right.\nn\\
&&\qquad\quad\left.-7 (P^2+Q^2)^2\right)\alpha
 + 8  \left(P^2-Q^2\right) \left( \left(3 P^2+Q^2-2 r_{b,0}^2\right)+6 r_{b,0}^4\ell^{-2}\right)\g_1\cr
&&\qquad\quad-32  P^2 \left(\left(P^2-r_{b,0}^2\right)+3 r_{b,0}^4\ell^{-2}\right)\g_2
\Big]\,, \nn\w2
&&\delta \psi^m_b=\frac{P}{20g\sqrt{\pi^3 G} r^8_{b,0}T_{b,0}}\Big[
\left(\ell ^{-2} \left(6 P^2+54 Q^2-5 \ell ^2\right) r_{b,0}^4+2  \left( Q^2- P^2\right) r_{b,0}^2+45 r_{b,0}^8
\right.\nn\\
&&\qquad\quad\left.+ (P^2+Q^2)^2\right)\alpha
 + 8  \left(P^2-Q^2\right) \left( \left(P^2+3Q^2-2 r_{b,0}^2\right)+6 r_{b,0}^4\ell^{-2}\right)\g_1\cr
&&\qquad\quad+32  Q^2 \left(\left(Q^2-r_{b,0}^2\right)+3 r_{b,0}^4\ell^{-2}\right)\g_2
\Big]\,, \nn\w2
&&\delta V_{b}=\frac{4}{5 G r^4_{b,0}T_{b,0}}\Big[
\left(5 (P^2-Q^2) r_{b,0}^2-2 (P^2-3 Q^2)(P^2+Q^2)\right)\alpha+4 (P^2-Q^2)^2\g_1\cr
&&\qquad\quad+16   P^2 Q^2\g_2\Big]\,,
\eea

In the modified Smarr relation \eqref{msmarr}, quantities proportional to the dimensional parameters $\a,\,\g_1,\,\g_2$ are given by
\bea
\Xi&=&\frac{ (r_{b,0}-r_{c,0})}{5 G r_{c,0}^5 r_{b,0}^5 (r_{c,0} r_{b,0}+r_{c,0}^2+r_{b,0}^2)}\Big[(P^2+Q^2) (P^2+7 Q^2) r_{b,0}^5 (2 r_{c,0}+r_{b,0})\cr
&& +2 r_{c,0}^3 r_{b,0}^3 \left(5 (P^2-Q^2) r_{b,0}^2-6 (P^2-3 Q^2) (P^2+Q^2)\right)\cr
&& +r_{c,0}^5 (r_{c,0}+2 r_{b,0}) \left(5 (P^2-Q^2) r_{b,0}^2
+(P^2+Q^2) (P^2+7 Q^2)\right)
\cr
&& +r_{c,0}^2 r_{b,0}^4 \left(5 (P^2-Q^2) r_{b,0}^2+3 (P^2+Q^2)(P^2+7 Q^2)\right)\cr
&& +3 r_{c,0}^4 r_{b,0}^2 \left(10 (P^2-Q^2) r_{b,0}^2+(P^2+Q^2)(P^2+7 Q^2)\right)
\Big]\,,\nn\w2
\Upsilon_1&=&-\frac{8 (P^2-Q^2)^2 (r_{c,0}^5-r_{b,0}^5)}{5   Gr_{c,0}^5 r_{b,0}^5},\quad \Upsilon_2=-\frac{32 P^2 Q^2  (r_{c,0}^5-r_{b,0}^5)}{5 G r_{c,0}^5 r_{b,0}^5}\,.
\eea

\end{document}